\documentclass[10pt]{iopart}
\usepackage{iopams}
\usepackage{graphicx}
\begin{document}

\title[Exact canonical occupation numbers in a Fermi gas]{Exact canonical occupation numbers in a Fermi gas with finite level spacing
and a $q$-analog of Fermi-Dirac distribution}
\author{Vyacheslavs Kashcheyevs}
\address{Faculty of Computing and Faculty of Physics and Mathematics,\\ University of Latvia, Riga LV-1586, Latvia}
\ead{slava@latnet.lv}
\date{October 25, 2011}
\begin{abstract}
We consider equilibrium level occupation numbers in a Fermi gas with a fixed number of particles, $n$, and finite level spacing.
Using the method of generating functions and the cumulant expansion we derive a recurrence relation for canonical partition function and
an explicit formula for occupation numbers in terms of single-particle partition function at $n$ different temperatures.
We apply this result to a model with equidistant non-degenerate spectrum and obtain close-form expressions
in terms of $q$-polynomials and Rogers-Ramanujan partial theta function.
Deviations from the standard Fermi-Dirac distribution can be interpreted in terms of a gap in the chemical potential
between the particle and the hole excitations with additional correlations at temperatures comparable to the level spacing.
\end{abstract}

\maketitle

\section{Introduction}
Applications of statistical mechanics
to fermion systems with discrete spectrum, such as semiconductor quantum dots~\cite{kouwenhoven2001,Alhassid00},
naturally involve
single-particle averages in statistical ensembles with a fixed
number of particles.
In particular, the kinetic theory of tunneling \cite{beenakker1PBI,AverinKorotkov1991}
through quantum dots with fast intra-dot electron relaxation involves
average level occupation number,
\begin{equation}\label{eq:avdef}
    \langle \nu_k \rangle_n = \frac{-1}{\beta Z_n} \frac{\partial Z_n}{\partial \epsilon_k} \, ,
\end{equation}
in a  Gibbs distribution of $n$ independent fermions populating
a set of single-particle energy levels $\{ \epsilon_k \}$ (enumerated by $k=0,1,2 \ldots$).
Here
$Z_n$ is the canonical partition function
\begin{equation} \label{eq:Zdef}
  Z_n = \sum_{ \{ \nu_k \} }\exp \Bigl ( -\beta \sum_k \nu_k \epsilon_k \Bigr ) \delta_{n, \sum_k \nu_k} \, ,
\end{equation}
and $\beta$
is the inverse thermodynamic temperature. For fermions, occupation numbers $\nu_k$ in the sum (\ref{eq:Zdef}) take values
$0$ and $1$.

The behavior of $ \langle \nu_k \rangle_n$ is simple in two extreme limits of the typical level  $\Delta$.
For $\beta \Delta \ll 1$ and large $n$,
Eqs.~(\ref{eq:avdef})-(\ref{eq:Zdef}) reduce to the standard  Fermi-Dirac distribution,
\begin{equation} \label{eq:FD}
  \langle \nu_k \rangle_n \to f(\epsilon_k - \mu) =\frac{1}{1+ e^{\beta (\epsilon_k -\mu)}} \, .
\end{equation}
Here $\mu$ is the chemical potential determined by the normalization
condition $n =\sum_k f(\epsilon_k - \mu)$.
In the low temperature limit,  $\beta \Delta \gg 1$, most of the statistical weight in Eq.~(\ref{eq:Zdef})
is in the ground state (defined as $\nu_k =1$ for $0 \le k <n $  and $\nu_k=0$ otherwise). In this case it is common \cite{Buttiker1985a,Buttiker1987}
to take only one excited state into account resulting in a two-state Gibbs distribution
which is equivalent to  Eq.~(\ref{eq:FD}) for $k=n\!-\!1,n$
with $\mu = (\epsilon_{n}\!+\!\epsilon_{n-1})/2$ and $\beta \to \beta^{\ast}=2 \beta$ ~\cite{beenakker1PBI}.

For finite $\beta \Delta $, the average
occupation number $  \langle \nu_k \rangle_n$ deviates from Eq.~(\ref{eq:FD}) in a non-universal way which depends on
the details of the energy spectrum \cite{beenakker1PBI,Alhassid1998,Zianni2007}.
Exact analytical investigation of this regime is complicated by the combinatorial explosion in the number of levels with comparable statistical weights.
Particle number projection technique based on Fourier extraction \cite{Ormand1994,Alhassid1998}
 gives an exact closed-form formula for
$\langle \nu_k \rangle_n$ (see Eq.~(140) of \cite{Alhassid00})
that scales quadratically with the number of levels; 
however, its potential for analytical investigation appears to be limited due to the additional sum over the Fourier variable.

In the paper we address the problem of exact evaluation of canonical occupation numbers  $\langle \nu_k \rangle_n$
by deriving a general formula that scales linearly in the number of levels and quadratically
in the number of particles, Eqs.~(\ref{eq:Znmain}) and (\ref{eq:occumain}) below.
The energy spectrum $\{ \epsilon_{k} \}$ enters the formula only in terms
of the single-particle partition function $Z_1$ computed at $n$ different temperatures.
We apply this general result to equidistant spectrum, $\epsilon_k = k \, \Delta $,
and derive an exact formula for $\langle \nu_k \rangle_n$  in terms of polynomials in $q \equiv e^{-\beta \Delta}$, Eq.~(\ref{eq:exactgFD}).
In the limit of degenerate Fermi gas, $n \beta \Delta \gg 1$, these polynomials converge to partial theta function \cite{andrews2008ramanujan}
which is involved in a number of combinatorial proofs \cite{Alladi2007,BerndtKimYee2010,Yee2010} of Ramanujan's identities \cite{RamanujanLost}.
The exact result for the equidistant spectrum can be approximated well
by tailoring
\emph{two} standard Fermi-Dirac distributions (\ref{eq:FD}) with different
chemical potentials for holes and for particles, $\mu_h = \epsilon_n$ and $\mu_p =\epsilon_{n-1}$, respectively.
In the high- and the low-temperature limits, this approximation converges to the asymptotically exact Fermi-Dirac and two-state Gibbs distributions,
respectively. At intermediate temperatures, $\beta \Delta \sim 1$,  particle-hole correlation effects
due to fixed $n$ result in finite deviations from the exact solution.

\section{General expressions for  fermion partition functions and the occupation numbers}
Grand canonical  partition function
$Y$ serves as a generating function for the canonical partition functions $Z_n$ if expanded
power series of $z = e^{\beta \mu_0}$,
\begin{equation} \label{eq:plainY}
  Y(z) = \sum_{ \{ \nu_k \} }\exp \Bigl ( -\beta \sum_k \nu_k (\epsilon_k-\mu_0) \Bigr ) = 1 +\sum_{n=1}^{\infty} Z_n z^n \, .
\end{equation}
$Y(z)$ is most conveniently calculated via its logarithm \cite{LL5},
 \begin{eqnarray} \label{eq:logY}
  \ln Y(z)  & =  & \sum_k \ln \left ( 1 + z e^{-\beta \epsilon_k} \right ) = \sum_{n=1}^{\infty} \frac{\kappa_n}{n!} z^n \, ,
 \end{eqnarray}
where
$\kappa_n \equiv (-1)^{n+1} (n-1)! Z_1(\beta n)$,
and
\begin{equation}
   Z_1(\beta') = \sum_k e^{-\beta' \epsilon_k}
\end{equation}
is the canonical partition function of a \emph{single} particle.

Relation between $n! Z_n$ and $\kappa_n$ is the same as between the raw moments and the cumulants 
of a univariate probability distribution and is given by the complete Bell polynomials \cite{Bell1934},
\begin{equation} \label{eq:Bell}
  Z_n =(n!)^{-1} B (\kappa_1, \kappa_2, \ldots, \kappa_n) \, .
\end{equation}
The latter  satisfy a recurrence relation \cite{Smith1995},
\begin{equation}
    B (\kappa_1, \kappa_2, \ldots, \kappa_n)   = \kappa_n +\sum_{m=1}^{n-1} { {n-1} \choose{m-1} } \kappa_m  B (\kappa_1, \kappa_2, \ldots, \kappa_{n-m})  \, ,
\end{equation}
which translates into
\begin{equation} \label{eq:Znmain}
       Z_n  =\frac{1}{n}  \sum_{m=1}^{n} (-1)^{m+1} Z_1(\beta m) Z_{n-m}  \, .
\end{equation}
We set $Z_0 = 1$ identically.

Combining Eqs.~(\ref{eq:avdef}), (\ref{eq:plainY}) and (\ref{eq:logY}) gives the generating function for the occupation numbers:
\begin{equation}
\sum_{n=1}^{\infty}  \langle \nu_k \rangle_n Z_n  z^n  = Y(z) \frac{z e^{-\beta \epsilon_k}}{1+z e^{-\beta \epsilon_k}} \, .
\end{equation}
Expanding the r.h.s.\ in powers series in  $z$ gives
\begin{equation} \label{eq:occumain}
   \langle \nu_k \rangle_n  = \frac{1}{Z_n} \sum_{m=1}^{n} (-1)^{m+1} e^{- \beta m \epsilon_k}  Z_{n-m} \, .
\end{equation}
Equations (\ref{eq:Znmain}) and (\ref{eq:occumain}) constitute our main general result.

\section{Example: equidistant spectrum}
\subsection{Exact finite-$n$ results: polynomials}
For $\epsilon_k = k\, \Delta$, the grand canonical partition function (\ref{eq:logY}) can be
expressed by an infinite product,
\begin{equation}
   Y(z) =\prod_{k=0}^{\infty}(1+q^k z)  = (-z; q)_{\infty} \, ,
\end{equation}
where  $q \equiv e^{-\beta \Delta}$ and $(\cdot ; q)_n$ is the  $q$-shifted factorial \cite{Gasper1990,DLMF}.

Using $q$-analog binomial theorem of Euler  (\cite{DLMF}, formula 17.2.35)
we can get the partition function directly
from Eq.~(\ref{eq:plainY}),
\begin{equation}
Z_n = \frac{q^{n(n-1)/2}}{(q; q)_{n} } \, .
\end{equation}

Applying Eq.~(\ref{eq:occumain}), and transforming $q$-shifted factorials (\cite{DLMF}, formula 17.2.13),
one gets
\begin{eqnarray} \label{eq:exactgFD}
     \langle \nu_k \rangle_n   & = & 1- p(k,n; q) \, \label{eq:gFDpoly1} \, , \\
     p(k,n; q) & \equiv  & \sum_{m=0}^{n} q^{m(k+1)} (q^{-n}; q)_m  \\
      & = &
      1+\sum_{m=1}^{n} \prod_{l=0}^{m-1} (q^{k+1} -q^{l+k-n+1}) \, . \label{eq:pdef}
\end{eqnarray}
Equation (\ref{eq:gFDpoly1}) defines occupation numbers for $n$ fermions populating equidistant levels at equilibrium.
It is clear from the explicit form (\ref{eq:pdef}) that  $p(k,n;q)$ is a Laurent polynomial
(the product contains negative powers of $q$ if $k<n$). However, since $0 \le \langle \nu_k \rangle_n \le 1$
for $q \to 0$ by definition (\ref{eq:avdef}), the negative powers of $q$ must cancel, thus we conclude that
$p(k,n;q)$ is  always an ordinary polynomial in $q$ for $n>0$, $k\ge 0$.
This cancelation is not trivial and deems further mathematical investigation~\cite{MathSE73536}.

A number of recurrence formulas can be derived for $p(k,n;q)$ \cite{Smotrovs2011},
including a symmetry relation
\begin{equation} \label{eq:symmetry}
  q^k p(k,n;q) =q^{n}  p(n,k;q)   \, .
\end{equation}
Using (\ref{eq:pdef}) in the r.h.s. of (\ref{eq:symmetry}) gives a sum of products with no negative powers of $q$ at $k<n$.

\subsection{Large-$n$ limit: $q$-analog of Fermi-Dirac distribution}
If $ \epsilon_n \gg \beta^{-1} $ then the Fermi gas is degenerate \cite{LL5} and the limit of $q^{n} \to 0$
is appropriate.
For $k, n\to \infty$, and $k-n =\mathrm{const} \ge -1$ the polynomial sum in Eq.~(\ref{eq:pdef}) becomes a geometric series which gives
\begin{eqnarray}
\lim_{{{n \to \infty}\atop{k\ge n-1}}}
p(k,n;q) & = & \theta(-q^{k-n+1/2},q^{1/2}) \, ,
\end{eqnarray}
where
\begin{eqnarray}
    \theta(a,q) & \equiv & \sum_{m=0}^{\infty} a^m   q^{m^2}
\end{eqnarray}
is known as partial theta function\footnote{
Note that $\lim_{n\to\infty} p(n,n;q) =\sum_{m=0}^{\infty} (-1)^m q^{m(m+1)/2} $ is an instance of \emph{false theta series} \cite{AndrewsWarnaar2007}
in the sense of L~.J.~Rogrers~\cite{Rogers1917}.}
\cite{andrews2008ramanujan}.
 The partial theta function is famous for a number of identities
discovered by Ramanujan in his lost notebook \cite{RamanujanLost}. These identities have been studied extensively \cite{
AndrewsWarnaar2007,andrews2008ramanujan}, including some recent proofs by combinatorial methods \cite{Alladi2007,BerndtKimYee2010,Yee2010}.
Using the symmetry relation (\ref{eq:symmetry}) for
$k<n$, gives the particle-hole complementary result : $\lim_{n\to \infty} p(k,n;q) = 1-\theta(-q^{n-k-1/2},q^{1/2}) $.

In terms of level energies $\epsilon_k$, the occupation numbers in a canonical degenerate Fermi gas with constant
levels spacing $\Delta$ can be written as
\begin{equation}
    \lim_{n \to \infty} \langle \nu_k \rangle_n  =
      \biggl\{
     \begin{array}{ll}
  f_q (\epsilon_k\!-\!\mu\!+\!\Delta/2 ) ,  &  \epsilon_k > \mu \!+\!\Delta/2  \, ,  \\
  1-f_q  (\!-\epsilon_k\!+\!\mu\!+\!\Delta/2 ) ,  &   \, \epsilon_k < \mu\!-\!\Delta/2  \, ,
    \end{array}  \label{eq:piecwiseform}
\end{equation}
where
\begin{eqnarray} \label{eq:qFDdef}
   f_q (\epsilon) = \theta( -e^{-\beta \epsilon}, q^{1/2}) \stackrel{q\to 1}{=} f(\epsilon) \, .
\end{eqnarray}
and $\mu=(\epsilon_{n}+\epsilon_{n-1})/2$.
The function defined in Eq.~(\ref{eq:piecwiseform}) can be considered a $q$-analog \cite{Gasper1990}
of the  standard Fermi-Dirac distribution (\ref{eq:FD}) since $\lim_{q\to 1} f_q(\epsilon) = f(\epsilon)$.
\begin{figure}
\begin{center}
  \includegraphics[width=0.7\textwidth]{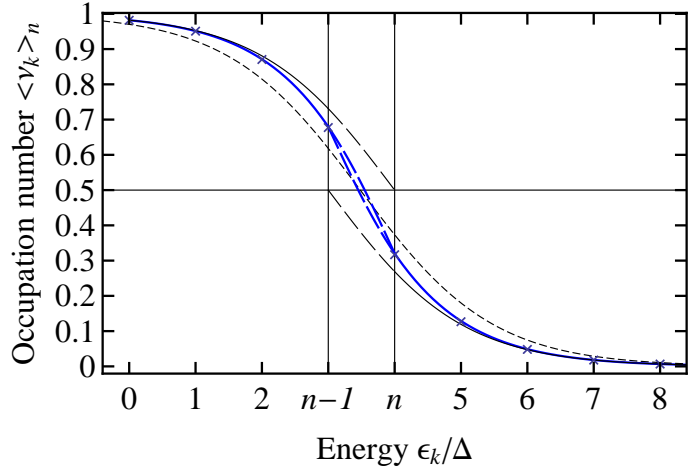}
\end{center}
\caption{Comparison of different approximations for the canonical occupation numbers in a Fermi gas with equidistant spectrum,
$n=4$ and $\beta \Delta =1$.
Crosses mark the exact values, thick (blue) continuous line shows the $q$-analog of Fermi-Dirac distribution, Eq.~(\ref{eq:piecwiseform}), thin continuous line --- substitution $f_q\!\to\!f$ in Eq.~(\ref{eq:piecwiseform}), and short-dashed line --- single Fermi-Dirac distribution, Eq.~(\ref{eq:FD}). The long-dashed lines between $\epsilon_{n-1}$ and $\epsilon_{n}$ are extrapolations of the corresponding functions into the gap $\mu_p < \epsilon  < \mu_h$.
\label{fig:fig1}}
\end{figure}

Equation (\ref{eq:piecwiseform}) expresses two essential deviations of canonical occupation numbers from Fermi-Dirac distribution.
Firstly,  the $q$-analog $f_q$ is different from $f$.
An approximation of substituting $f_q\!\to\!f$ in Eq.~(\ref{eq:piecwiseform}) becomes exact both for $q\!\to\!1$ and for $q\!\to\!0$.
Numerically, we find
maximal absolute deviation $|f(\epsilon_k)-f_q(\epsilon_k)|$ of  $0.0567$ reached for $k=n,n\!-\!1$ at $\beta \Delta=0.752$.
For the two levels closest to the gap, $k=n\!-\!1,n$ approximating $f_q\!\to\!f$ gives $\langle {\nu_{n}}
\rangle_n =1-\langle{\nu_{n-1}}\rangle_n=
\exp(-\beta \Delta/2)/ [ 2 \cosh (\beta \Delta/2 )]$ which is equivalent to two-state Gibbs
approximation \cite{beenakker1PBI}.
A comparison between the exact result (\ref{eq:exactgFD}), the large-$n$ limit (\ref{eq:piecwiseform})  and
a single Fermi-Dirac distribution is shown in Fig.~\ref{fig:fig1} for $n=4$ and $\beta \Delta =1$.

Secondly, barring the difference between $f_q$ and $f$, Eq.~(\ref{eq:piecwiseform}) can be seen as a combination
of two Fermi-Dirac distributions with different chemical potentials for particles, $\mu_p =\mu\!-\!\Delta/2 =\epsilon_{n-1} $, and
for holes, $\mu_h =\mu\!+\!\Delta /2=\epsilon_{n}$, respectively. This is precisely what is
to be expected if one considers particle and hole excitations from the ground state as two uncorrelated microcanonical ensembles.
In this view, the moderate difference in the functional dependence between $f(\epsilon)$ and $f_q(\epsilon)$
results from correlation between a particle and a hole created in a single pair-excitation act.
Note that the difference $\mu_h-\mu_p=\Delta $ can not be ignored even as $\beta \to \infty$ and $q\to 0$,
thus using a single Fermi-Dirac distribution, Eq.~(\ref{eq:FD}), necessarily fails for finite $\Delta$
and low temperature.

\ack
The author is grateful to Carlo Beenakker, Bruce C. Berndt, Jan Mangaldan, Juris Smotrovs, and Michael Somos for discussions.
This research has been supported by ESF project no. 2009/0216/1DP/1.1.1.2.0/09/APIA/VIAA/044.

\section*{References}

\providecommand{\newblock}{}

\end{document}